# Analyzing 3D Volume Segmentation by Low-level Perceptual Cues, High-level Cognitive Tasks, and Decision-making Processes

Anahita Sanandaji[1], Cindy Grimm[2], Ruth West[3], Max Parola[3], Meghan Kajihara[3], Kathryn Hays[3], Luke Hillard[3], Brandon Lane[3], and Molly Beyer[3]

*[1]Ohio University, [2]Oregon State University, [3]University of North Texas*

**Abstract**
3D volume segmentation is a fundamental task in many scientific and medical applications. Producing accurate segmentations efficiently is challenging, in part due to low imaging data quality (e.g., noise and low image resolution) and ambiguity in the data that can only be resolved with higher-level knowledge of the structure. Automatic algorithms do exist, but there are many use cases where they fail. The gold standard is still manual segmentation or review. Unfortunately, even for an expert, manual segmentation is laborious, time consuming, and prone to errors. Existing 3D segmentation tools are often designed based on the underlying algorithm, and do not take into account human mental models, their lower-level perception abilities, and higher-level cognitive tasks. Our goal is to analyze manual segmentation using the critical decision method (CDM) in order to gain a better understanding of the low-level (perceptual and marking) actions and higher-level decision-making processes that segmenters use. A key challenge we faced is that decision-making consists of an accumulated set of low-level visual-spatial decisions that are inter-related and difficult to articulate verbally. To address this, we developed a novel hybrid protocol which integrates CDM with eye-tracking, observation, and targeted questions. In this paper, we develop and validate data coding schemes for this hybrid data set that discern segmenters' low-level actions, higher-level cognitive tasks, overall task structures, and decision-making processes. We successfully detect the visual processing changes based on tasks sequences and micro decisions reflected in the eye-gaze data and identified different segmentation decision strategies utilized by the segmenters.

# 1. Introduction

3D volume segmentation is a fundamental step in many applications such as biomedical imaging (e.g. locating tumors, measuring tissue volumes and computer guided surgery) and scientific study (e.g., understanding how $CO_2$ flows through rocks). Creating 3D surface structures from segmentations is often required to understand the 3D image content (e.g., radiation treatment volumes, airflow simulations of rabbit nasal passages).

Because there are many situations where automatic algorithms fail, we focus on manual segmentation and review as a gold standard for creating 3D structures. This is a non-trivial and time intensive task that it involves making many local decisions based on image features (e.g., as in Figure 1 boundaries of biological tissues are not always clearly separated) and potentially spatially and topologically complex 3D structures. Low image resolution and noise can also make segmentation challenging, particularly for novice segmenters. In addition, the segmentation is usually performed (or evaluated) on 2D slices of the 3D data and the segmenter must mentally integrate these into a coherent 3D structure. As a result, 3D segmentation of biomedical data is usually done by experts who have many years of experience in segmenting specific anatomical structures. In summary, experts create 3D segmentations using low-level perceptual cues to make low-level decisions on marking, all guided by higher-level cognitive tasks. Visual cues include intensity, spatial, and textural attributes (Soh & Tsatsoulis, 2000), low-level marking tasks include delineating structures in an image plane by marking contours or filling regions. Experts also use higher-level constraints — such as connectivity, shape and topology — to disambiguate unclear structure boundaries (Ju, Zhou, & Hu, 2007).

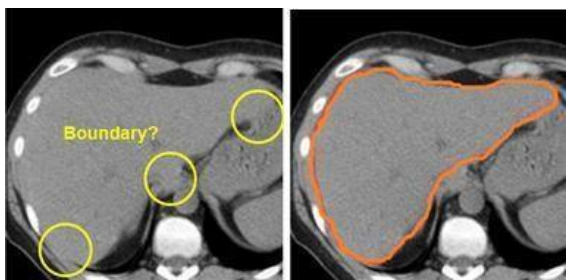

Figure. 1. Liver data set (left) and its contour (right). The boundaries (of the yellow circles) are not clearly separated in the left image.

Despite the large amount of work on developing interactive/semi-automated segmentation tools, there is little research on the correct way to design, and evaluate, such systems, nor is there research on how to capture, in a fine-grained way, what experts are doing in the 3D volume segmentation, and what their key decision-making elements are.

Our goal is to capture both low-level decisions and high-level segmentation actions and tasks performed by segmenters as a necessary preliminary step before defining more hypothesis-driven studies. It may seem obvious that segmenters look at what they are segmenting, or that they explicitly can tell us what their key decisions are. However, we do not know the specific image features experts use, how those might vary with the segmentation task, how experts determine their segmentation strategies and what visual decision-making process they have, and how that might vary during segmentation (which can often take hours).

To approach this problem, we began with field studies where researchers use their own tools on their own data sets because knowledge of the data set and tool capabilities are key to successful segmentation. Although standard Cognitive Task Analysis (CTA) (Clark, 2004, 2006; Clark, Feldon, van Merrienboer, Yates, & Early, 2008) or Critical Decision Method (CDM) (Klein, Calderwood, & Macgregor, 1989) approaches are useful for extracting higher-level task breakdowns (get the data, decide where to start segmenting and with what tool), in practice they are not capable of capturing sub-tasks (what to mark when), which is largely a visual-spatial decision-making process. In essence, the segmentation process consists of an accumulation of small, individual micro decisions made by visually examining the data. For this reason, we developed a hybrid field-study protocol that incorporates elements of CTA, CDM, and eye-tracking/marking data (West et al., 2016).

In this paper we describe our data coding scheme which lets us combine data from a more traditional CDM approach with our eye-tracking/marking data. This coding scheme lets us successfully detect the visual processing changes based on tasks sequences and micro decisions reflected in the eye-gaze data and identified different segmentation decision strategies utilized by the segmenters. In specific, we use this coding scheme to answer the following research questions:

- RQ1: Can we identify where segmenters look and what their low-level (micro) decisions and higher-level (macro) segmentation actions and tasks are?
- RQ2: How do expert and novice segmenters' gaze patterns, and therefore micro decisions, differ for different/similar segmentation micro/macro tasks?

Contributions of our work include:
1. Integrating critical decision method (CDM) with eye-tracking to elicit visual micro-decisions of segmenters.
2. Devising a novel coding scheme that encodes segmenters' observable actions and micro-decisions, and what they are looking at when performing those actions.
3. Creating a domain-agnostic classification to capture segmentation-level tasks.
4. Defining different segmentation approaches and decisions strategies used by participants.
5. Although we focus here on one application (segmentation), this general hybrid-study plus data coding method is potentially applicable to any task that has a substantial visual-spatial decision-making component.

## 2. Background

Substantial work has focused on creating and evaluating various 3D image segmentation methods but only a few have taken into account humans and their role. Olabarriaga & Smeulders (2001) presents an early review of human-computer interaction in image segmentation, mentioning three common criteria for segmentation evaluation: Accuracy, repeatability and efficiency. In (Krupinski, 2010), the author aims to understand how physicians interact with the information in a medical image during the interpretation process. However, medical imaging perception and interpretation is only a part of the 3D volumetric data segmentation process.

In order to assess the quality of expertise differences in the comprehension of medical visualizations, Gegenfurtner, Siewiorek, Lehtinen, & Saljo (2013) reviews quantitative studies that characterize where experts look based on where in the image their gaze rests (fixation) and how long it is there (dwell time). Fixation and dwell time are lower level tasks that relate to higher order tasks of searching, abnormality detection and accuracy of detection. Segmentation, while requiring the ability to detect a feature of interest, also requires the expert to build a 3D structure from those features. Recent work by (Li, Pelz, Shi, Alm, & Haake, 2012) lays the groundwork for understanding what effective visual strategies experts employ for examining medical images. Their work can be used for advanced medical image understanding; however, it does not cover 3D volume segmentation.

The authors of (Ramkumar et al., 2016) investigate effects of user interaction in semi-automatic segmentation methods for segmenting the organs at risk in radiotherapy planning. Their findings suggest that in the future HCI design of semi-automated segmentation approaches, there should be more flexibility in the interface design, and user interactions should be less cognitively challenging. Their work reflects the mode of data capture, the logic of underlying algorithms, and input devices with the aim of providing flexibility in user interfaces or decreasing the cognitive load of interactions within the overall process. Unfortunately, they only had two participants in their study which limits inter-observer variations. The focus of HCI optimization ultimately arises from the concept of fewer clicks, translating to ease of use rather than from the notion of domain knowledge as supportive of ease of use.

In our previous publication (Sanandaji, Grimm, West, & Parola, 2016), we focused on capturing experts' gaze location during 3D volume segmentation. Now, for this paper, we report findings for more participants (both novice and experts), not only covering where segmenters look, but also identity their low/higher level actions, tasks, and strategies during 3D volume segmentation.

## 3. Field Study Experiment

Our research goal is to analyze manual segmentation using the critical decision method (CDM) in order to understand human factors that are involved in a 3D volume segmentation process. We aim to capture enough domain knowledge in order to effectively help segmenters, but not to become experts in segmentation for any particular domain (Rogers, 2004; van Wijk, 2006). Understanding 3D volume segmentation process, from a human-factors perspective, is challenging and requires enough attention to both study design and implementation. Through informal interaction with segmentation experts and novices, we developed an initial understanding of the segmentation process. Based on that, we define 3D volume segmentation to be an iterative process that incorporates several steps including inspection, navigation, marking, editing, and visualization. Each of these steps requires various sets of sub-tasks and has its own constraints and challenges.

We conducted formative field studies to further understand the segmentation process in the context of experts performing real segmentation tasks, and to capture both low-level (micro) and higher-level (macro) decisions during 3D volume segmentation. We chose field studies and in-depth, per-expert analysis over specific controlled testing because the tools, data sets and expert approaches are very idiosyncratic, and running a formalized controlled study too early risks missing non-obvious elements or biasing our results.

### 3.1 Participants

We have 10 participants from 4 different sites recruited through personal contacts. Tools used ranged from custom-built to commercially available software. These participants cover a range of tasks, data sets, tools, and expertise. Because segmentation is a complex task that involves different visual search skills, decision-making elements, and domain knowledge, we believe that using a small number of experts that are well-trained in the respective areas of the tasks is warranted (Biederman & Shiffrar, 1987; Cohen & Hegarty, 2007). We balance the small number of experts in each domain by recruiting participants with several diverse domains, software and segmentation tools. Our participants are representative of segmenters who are working on 3D volume segmentation projects. Table 1 summarizes the domain and expertise of the participants of our field studies. Images of the datasets our participants worked with are shown in Figure 2.

Table 1. Participants of the study.

| Participant | P1 | P2 | P3 | P4 | P5 | P6 | P7 | P8 | P9 | P10 |
|---|---|---|---|---|---|---|---|---|---|---|
| Expertise | Radiology | Biomedical | Soil Analysis | Cell Biology | Cell Biology | Cell Biology | Cell Biology | Cell Biology | Cell Biology | Cell Biology |
| Years Segmenting | >5 yrs. | >3 yrs. | >1 yr. | <1 yr. | <1 yr. | <1 yr. | >5 yrs. | >1 yr. | >3 yrs. | >40 yrs |
| Prior Experience with Data | No | Yes | Yes | Yes | Yes | Yes | Yes | Yes | Yes | Yes |
| Site | Site 1 | Site 1 | Site 2 | Site 2 | Site 3 | Site 3 | Site3 | Site 4 | Site 4 | Site 4 |
| Tool | Tool 1 | Tool 1 | Tool 2 | Tool 3 | Tool 3 | Tool 3 | Tool 3 | Tool 4 | Tool 4 | Tool 4 |

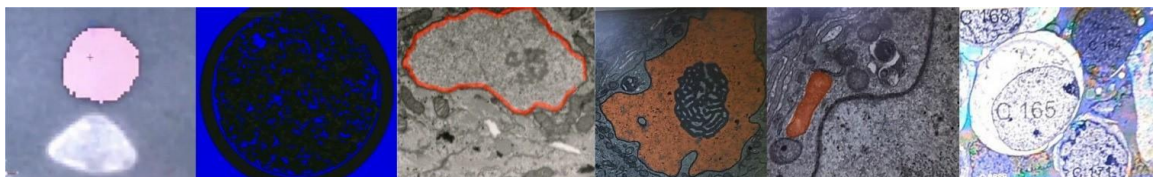

Figure 2. Data set examples. Left two: $CO_2$/soil analysis, right four: cell analysis.

### 3.2 Data and Procedure

To replicate segmenters' normal process as much as possible, we need to observe them in their work settings, using their own tools, data, and computers. The aim is to not only understand the segmentation process, but also how novices differ from expert segmenters. Therefore, we chose a field study (Yin, 2013), with modifications to account for the segmentation (West et al. 2016).

Our study protocol is summarized in Figure 3 and involves elements of Over the Shoulders video (OTS), ethnographic Semi-Structured Interviews (SSI), Retrospective Think-aloud (RTA), Cognitive Task Analysis (CTA), Critical Decision Method (CDM), and Eye Movement Tracking (ET). We describe two study protocols; the initial study protocol was revised because we discovered after analysis that the think-aloud protocol disrupts normal eye-gaze. In addition, we decided to integrate CDM and CTA with eye-tracking to elicit from the segmenters the cues, strategies, and goals associated with each step of the segmentation process.

44

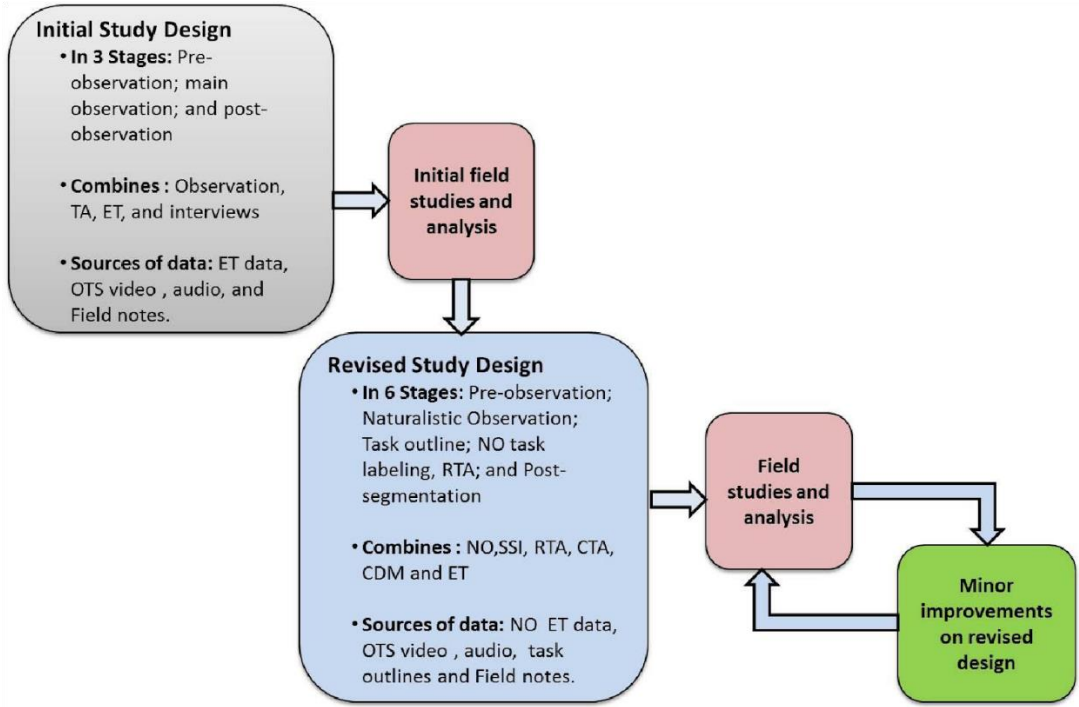

Figure. 3. Initial and revised design of our study protocol in a glance: Initial study involves Over the Shoulders video (OTS), Eye-tracking (ET), and Think-Aloud (TA); Revised study combines Naturalistic Observation (NO) with Semi-Structured Interviews (SSI), Retrospective Think-aloud (RTA), Cognitive Task Analysis (CTA), Critical Decision Method (CDM), and Eye-Tracking (ET).

### 3.2.1 Initial Study Design (P1-P4)

Through informal interaction with segmentation experts and novices we developed an initial understanding of the kinds of limitations and problems encountered in the field. This provided us with our initial design to study four participants (P1-P4). This design basically includes field observations and interviews, and is conducted in three stages of pre-, main, and post-observation. The eye-tracking (ET) data along with over the shoulder video (OTS) recording are the main sources of data.

In the initial design, the video was captured by placing a video camera over the shoulder of the participant in order to capture the screen and mouse. The eye-tracking utilized a 500 MHz under the monitor SMI tracker. We recorded audio both through the video camera, separate microphone, and the eye-tracker device (we merged the over the shoulder video with the eye-tracking data to achieve higher quality video), to make sure the video and eye-tracker data were synchronized. Study sessions lasted on average an hour, with a maximum of 2 hours. The on-site study team consisted of 3 researchers: A primary interviewer who asked the questions; a secondary interviewer who was responsible for quality control and follow up questioning; and a note taker who also captured video, sound and eye-tracking.

The pre-observation questions were warm-ups used to relax the participants and get an approximate idea of how and why the participants were segmenting the data. Because this stage did not involve working with the tool or performing any segmentation, we only recorded audio. Hand-written notes were also used for data triangulation. During the observation step, participants were asked to think-aloud while doing their segmentation tasks as close to their normal process as possible. For longer or repetitive segmentation tasks, participants demonstrated examples of how they performed different stages of the segmentation task. Finally, the post-observation questions covered high-level strategies and mental models, repeatability measures, post processing, and what necessitates redoing segmentation. Participants were free to use their segmentation tools and other downstream applications to answer these questions.

As cognition and perception are linked (Gegenfurtner et al., 2013), in studies where eye-tracking is required, the use of a think-aloud protocol while capturing eye movements can negatively affect the participant's eye motion and locus of attention during task performance. We realized that our participants also did not look at the place we expected them to look (e.g. the region of the structure) while thinking aloud. Therefore, we could not use all of the observation period data for our analysis. For example, for one participant we ended up only using 20 minutes of the one hour recorded eye-tracking data. For this reason, we switched to an uninterrupted segmentation recording, without the participant talking or interacting with the experimenters in any way. In our revised design, we address these limitations of the initial design.

### 3.2.2 Revised Study Design (P5-P10)

Our revised study design combines ethnographic Semi-Structured Interviews (SSI) with Retrospective Think-aloud (RTA) (Elling, Lentz, & de Jong, 2011; Hyrskykari, Ovaska, Majaranta, Räihä, & Lehtinen, 2008), Cognitive Task Analysis (CTA) (Clark 2004), Critical Decision Method (CDM) (Klein, Calderwood, & Macgregor, 1989), and Eye Movement Tracking (ET). CTA stopped well before the mark-off-a-contour level, so without eye-tracking video we could not have any information at that granularity. Replaying the eye-tracking video to the participants helped motivate them to talk more about their fine-grained decision-making process. Also, combining CDM with eye-tracking helped us elucidate both low-level sub-tasks, and higher level complex workflows and task sequences in the field. We used these task sequences to detect the visual processing changes and identify experts' strategies and key visual-spatial decisions during 3D volume segmentation process.

In (West et al., 2016) we presented a full description of this protocol. In this paper, we focus on the steps relevant to our analysis. We followed the basic pre-, main, and post-observation format used in the initial study, but substantially reduced the pre-observation stage and replaced the main Think-aloud stage with a Naturalistic Observation (NO) phase, followed by a Retrospective Think-aloud. In the NO phase, we asked the experts to perform their segmentation task without talking, and as close to normally as possible (again doing a few examples of repetitive tasks). We used eye-tracking (utilized a 500 MHz SMI remote glasses-based tracker) and video/audio capturing as sources of data. After the NO phase, we asked the participants to create a Task Outline (TO) of the segmentation process (on paper). Finally, during the post-observation phase, we asked them to review the NO video with us to identify the tasks, subtasks, and steps in their TO (a modified version of a retrospective think aloud). We used video/audio and handwritten notes as sources of data. Figure 4 shows an example of observation environment.

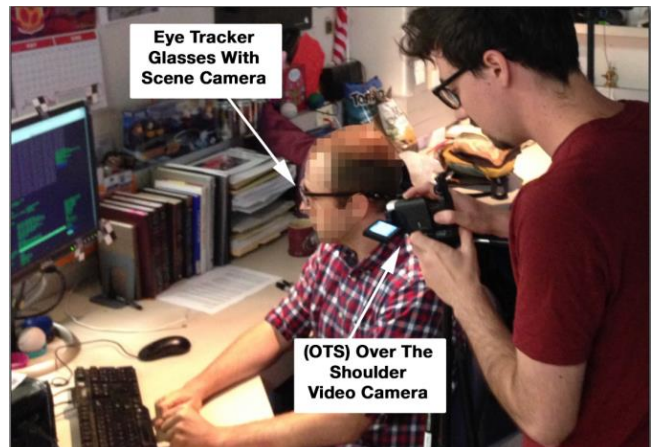

Figure. 4. Observation environment.

### 3.2.3 Eye-tracking Data Cleaning and Gaze Quality

We used the software supplied with the eye tracker to create the overlays. We constantly monitored the tracking during the study capture and re-did the calibration as necessary when it drifted, using times when gaze should match cursor location or button (e.g. adjusting a slider). While segmenting, the participants' gaze primarily consisted of smooth pursuit along identifiable features in the image data or transitions from a tool menu on the side to the data; therefore, we did not have to make judgments about rapid saccades within a small region or try to determine which of many small features the gaze overlapped.

### 3.2.4 Micro-Decision and Macro-Task Definition

We use different sources of data to capture two distinct categories of segmentation decisions and tasks, which we call micro-decisions and macro-tasks. Broadly speaking, micro-decisions are low-level visual-spatial actions that can be observed in the eye-tracking video, while macro-tasks come from the participant's cognitive task outlines. These are summarized in Table 2.

Table 2. Micro-decision and Macro-task Definition

| Category | Micro-decision | Macro-task |
|---|---|---|
| Definition | Visual spatial decision-making process by segmenters to accomplish low-level actions which are observable in an observation/NO video. We consider these actions something we can see in the video, not something we assume participants are doing. | Higher-level segmentation tasks which naturally consist of smaller sub-tasks and low-level actions. We identify these tasks through the CTA protocol and participants' task outlines |
| Goal | To identify where segmenters' are looking during segmentation, and to quantify lower-level actions and tasks of participants in a segmentation process. | To identify what participants are doing during different stages of the segmentation process, and what are the domain-agnostic phases and patterns of higher-level actions during segmentation. |
| Data Source | Observation (initial study) and NO (revised study) video (including eye-tracking video and OTS) | Combination of participants handwritten task-outline, task labeling, RTA and CDM |
| Example | Looking at the tool or image, paging, zooming, and drawing a contour | Data initialization, inspection, working with 2D and 3D view, and reviewing |

## 4. Data Processing and Analysis

The Naturalistic Observation (NO) eye-tracking data and scene camera video along with NO over the shoulder video serve as the foundation to capture micro-decisions. Macro-task capturing is accomplished using task-outline, task labeling, RTA, and CDM. We define our analysis process as following:

1. Combine NO videos with Heatmap/gaze overlays to determine participants' gaze location. We utilize these data to capture micro-decisions.
2. Create a novel coding scheme to categorize micro-decision elements. We assign a code to each identified micro-decisions. The coding scheme is in a hierarchical format.
3. Modify hand-written task-outlines to be simplified or extended. Then, create a mechanism to classify macro-tasks based on the modified task-outlines.
4. Micro-decisions analysis based on coding scheme to count the code frequency for each of the decision elements. This step validates our coding scheme from step 2.
5. Macro-task analysis using modified task-outlines to identify what participants are doing during different stages of the segmentation process.

### 4.1 Micro-decision Coding Scheme

The goal of developing a coding scheme is to support analysis and classification of participants' micro-decisions which involve perceptual cues, low-level actions (e.g. drawing) and tasks (e.g. marking vs. low-level reviewing). To our knowledge, the research literature does not report coding schemes that are directly applicable to 3D segmentation.

We consider a micro-decision to be something we can see in the video data rather than actions we think or assume the participant is doing. We associate a dedicated code to each visual spatial decision element. Micro-decisions naturally formed a hierarchy and are grouped into five higher-level action types: 1) Eye-gaze location: participant's eye-gaze location; 2) Global navigation: navigating between slices. 3) Local navigation: zooming and panning; 4) Marking: drawing or editing a contour; and 5) Review: reviewing a segmentation process or mark. Table 3 shows overview of our coding scheme. We now briefly describe our coding instructions, and how we used the coding scheme to capture micro-decisions.

**Eye-Gaze Location: Data and Tool:**

The primary objective of these codes is to identify when the eye-gaze is on the tool/UI and when it is on the data. To be considered "Data", the eye-tracker gaze overlay must be completely inside the data/image plane, or at a minimum it is touching the outside edge of the data/image plane. To be considered "Tool", the gaze overlay should completely be inside the tool UI, or at a minimum it is touching the outside edge of the tool UI. Figure 5 shows an example for "Data" and "Tool".

Table 3. Micro-decision Coding Scheme

| Action Type | Code | Description |
|---|---|---|
| **Eye gaze Location** | Data | User's eye gaze is completely inside the data/image plane itself or touching the edge. |
| | Tool | User's eye gaze is inside the tool UI, or at least touching the outside edge of the frame |
| | Boundary | User's eye gaze is in contact with the edge of the region of interest. |
| | Region | User's eye gaze is fully within the edge of the region of interest |
| | Non-ROI Features | Eye gaze in anywhere in the data image that is not the region of interest or its boundary. |
| | Movement | Eye gaze while it is in motion from Tool to Data |
| **Global Navigation** | Page | User navigates between slices/planes of the data. |
| | Free Rotate | User rotates data volume freely in x,y,z axes |
| **Local Navigation** | Pan | User drags view of data along x or y axis |
| | Zoom | User expands view over a portion of the data |
| **Marking Action** | Draw | User manually creates a mark by dragging their cursor, usually to encapsulate a figure or draw a line. |
| | Mark | User creates a special mark (e.g. circles) using the segmentation tool. |
| | Fill | A region of the data/image plane is filled in |
| | 1st Mark | Denotes the creation of the first mark in a data set and the processes leading up to it. |
| | Commit | A created mark is saved or approved as done by the user in the program |
| | Edit | Changes made to an existing mark |
| | Delete | Mark is removed |
| **Review** | Mark Review | Eye gaze lingers on a completed mark |
| | Assess Data | Eye gaze is on the data. Does not occur during marking action or when looking at a mark |
| | Copy | Propagation of the settings or marks from one slice to a different slice |
| | Tool Output | The output that the tool generates. |

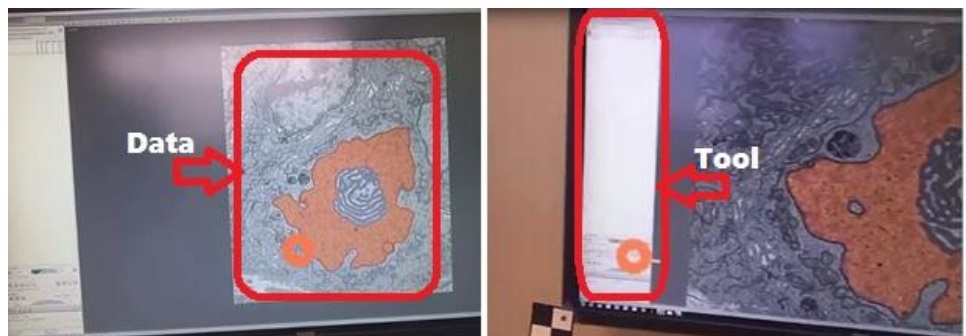

Figure. 5. Eye-gaze location examples for Data (left) and Tool (right). The small orange circle is the eye-gaze overlay.

**Eye-Gaze Location: Boundary, Region, Non-ROI features, and Movement:**

The primary objective of these codes is to identify where in the data the eye-gaze is pointing. These codes only occur within "Data" codes. "Region", "Boundary", and "Non-ROI" signify whether the gaze is within the region of interest, on the boundary of the region of interest, or somewhere else. "Movement" occurs when the eye-gaze shifts from the data to the tool/UI. To be considered "Boundary", the eye-tracker gaze overlay must be on the edge of the ROI or must be touching the edge either inside the ROI or outside. To be considered "Region", the gaze overlay should be inside the region but not touch the edge inside the ROI. Figure 6 shows examples for "Boundary" and "Region" codes. To be considered Non-ROI Feature, the gaze overlay is anywhere in the data that is not region or boundary (e.g. not touching the ROI and still within the data/image plane. "Movement" code captures when the eye-gaze is transitioning from one window (tool or data) to the other. This code is to differentiate between when the expert is actually looking at the data or tool and when their eyes are on a window because they are looking back and forth between windows/screens.

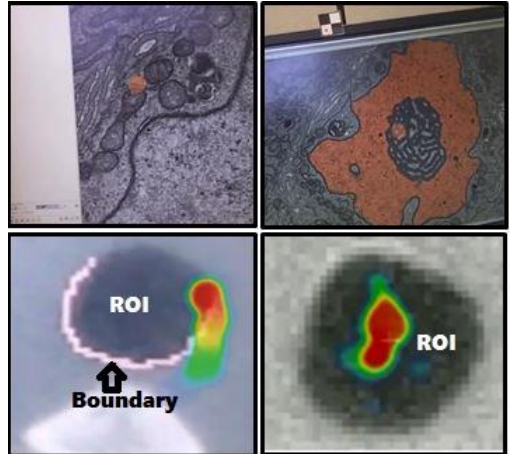

Figure. 6. Eye-gaze location examples for "Boundary" (top left and bottom left) and "Region" (top right and bottom right). The orange circle (in top images) and heatmaps (in bottom images) are the eye-gaze overlays.

**Mark and Navigation Codes:**

The objective for these codes is to identify marking and navigation micro-decision during segmentation. These codes may occur across both "Tool" and "Data" codes because some of these decisions are completed through interacting with the tool/UI. Navigation codes like "Page" are easier to identify in the NO video, so it is more efficient to initially annotate video using these codes and then continue with marking codes. Figure 7 and 8 show examples for "Draw" and "Fill" codes.

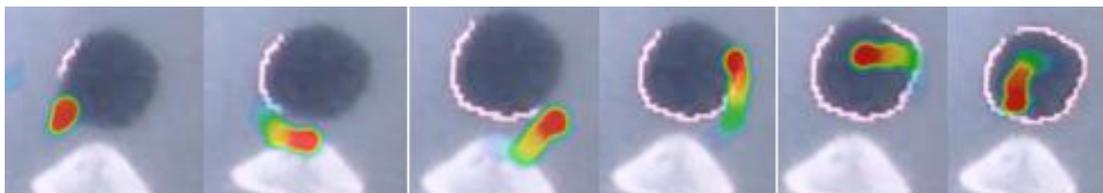

Figure. 7. "Draw" Code Example. Participant captures the structure by drawing (pink circle). Their heatmap gaze overlay follows the pink circle.

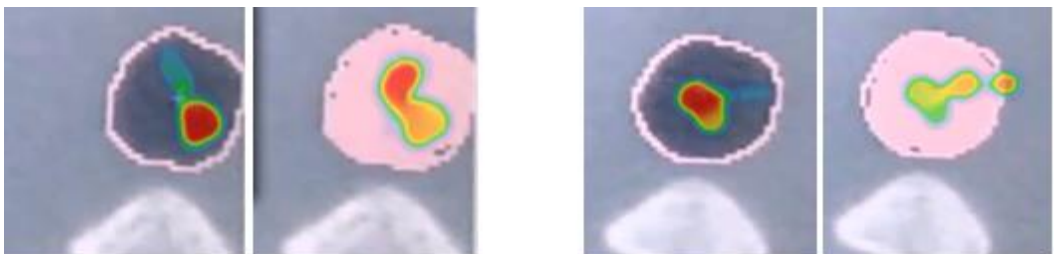

Figure. 8. "Fill" Code Example. Before and after Filling the structure. The gaze overlay is inside the region of interest.

**Review and Assessment Codes:**

The objective of these codes is to identify micro-decision attached to some segmentation thought and action processes. These codes belong under the "Data" code. "Assess Data" occurs when the eye-gaze is on the data, but a mark has not been made yet. "Review Mark" occurs when the eye-gaze is on a mark or commit that has been made. "Tool Output" occurs when the segmented slices are put together to create a 3D representation/model of a structure.

## 4.2 Macro-Task Classification

The goal of macro-tasks classification is to identify what participants are doing during different stages of the segmentation process (from the segmenters' perspective), what are the domain-agnostic phases and patterns of higher-level actions during segmentation, and what are their higher-level decision-making process. The input data stream is a combination of participants' handwritten task-outline, task labeling, RTA and CDM. We can only apply this coding to our revised studies (P5-P10), since we did not explicitly capture task outlines in the initial study. The participants' original task-outlines (Figure 9 a) had to be processed both to remove domain-specific comments and because they did not always match up exactly with the observed task sequences.

Task-outlines were not always complete. Sometimes participants were doing observable tasks in the NO videos, but do not talk about them in their task-outlines. If participants have missed anything, we use the NO videos to extend and complete the task-outlines. In addition, participants may talk about lots of things that were not necessarily relevant to their actual task or useful for analysis. Therefore, we remove those from the task-outlines as appropriate.

Using the modified task-outlines we group task items into higher-level macro-task categories. Table 4 shows our domain-agnostic macro-task categories. Figure 9 b) shows modified task-outline with macro-task categories for P5.

Table 4. Macro-task Classification

| Macro-Task | Description |
| --- | --- |
| **Setup** | Any task related to setting up, e.g., loading data, saving, and setting view |
| **Inspection** | Tasks related to inspecting ROIs, e.g., identification, looking, and scanning |
| **Tool Usage** | Using segmentation tools for marking structures (e.g., drawing) and function manipulation (e.g., changing threshold) |
| **Review** | Tasks related to reviewing segmentation results, including editing |
| **2D Navigation and View** | Moving through slices or using alternative 2D views |
| **3D Navigation and View** | Using 3D models, rendering and navigation |

a)

b)
1. Create new material [setup]
2. Annotate ROI [tool usage]
    1. Mark structure [tool usage]
        1. Adjust threshold of Magic Wand [tool usage]
        2. Grow selection [tool usage]
        3. Use Paint Brush tool [tool usage]
        4. Use Fill function [tool usage]
    2. Flip between slices [inspection] [2D navigating and view]
3. Interpolate [tool usage]
    1. Select the material of current slice and previous slice [tool usage]
    2. Page through a few slices [inspection] [2D navigating and view]
    3. Use Interpolate function [tool usage]
4. Review [review]
    1. Flip between slices [review][inspection] [2D navigating and view]
    2. Review annotation [review]

Figure. 9. a) Original task-outline. b) Modified task-outline with macro-task categories for the original one.

# 5. Results

We analyzed the coded data using the three following methods:

**1) Micro-decision frequency analysis**: Using the micro-decision coding scheme, we compute code frequency and average duration of each micro code interval over the entire observation. Micro-decision frequency analysis helps understand what low-level visual-spatial decisions participants take most often and how interleaved these decisions are.

**2) Macro-task frequency analysis: Using** the macro-task categories, we compute code frequency and average duration of each macro-task category interval over the entire observation. Macro-task frequency analysis helps understand what higher-level tasks participants take most often and how interleaved these tasks are.

**3) Snapshot analysis**: We analyze the data using video snapshots in order to distinguish how patterns change over time for different decisions/tasks. "Snapshots" are selected video segments corresponding to each of the macro codes and task items in modified task-outlines. For this analysis, we focus on code frequencies for eye-gaze location (specifically "Boundary" vs. "Region") during each of the macro-tasks. Snapshot analysis identified what participants are doing during different stages of the segmentation process.

## 5.1 Results of Micro-Decision Frequency Analysis

We have completed micro-decision frequency analysis for all 10 participants (P1-P10). For P7, we analyzed the two observations separately and in two sessions (Session 1 and Session 2). P7 was an expert participant who segmented two different datasets in two different NO sessions.

For this analysis, we first count the frequency of each micro-decision code over the entire valid observation/NO period. (e.g., if participant eye-gaze is on the tool for 26 times, we count 26 intervals, and the code frequency for the "Tool" micro-code is 26). We then capture average duration by taking the average time spent in each interval (e.g., if a participant has 26 tool intervals, totaling 58 seconds, the average duration is 58/26 = 2.2 seconds). We chose this approach because actions typically overlap so chunking up the video is not appropriate. Code frequency also lets us successfully capture micro-decisions and actions with very short durations. Now we present our results for each of the micro-tasks.

**Tool vs. Data Results:**

Figure 10 indicates that for each participant the frequency of looking at data is almost equal to looking at the tool, except in the case of participants of site 4 (P8-P10) who did not interact with the tool very much. The average duration spent looking at the data is greater than looking at the tool. The exception to this is P3, whose primary tool was a slider used to adjust a threshold value (no contour drawing). These results show that most participants go back and forth between the tool and data, but they spend more time gazing at data.

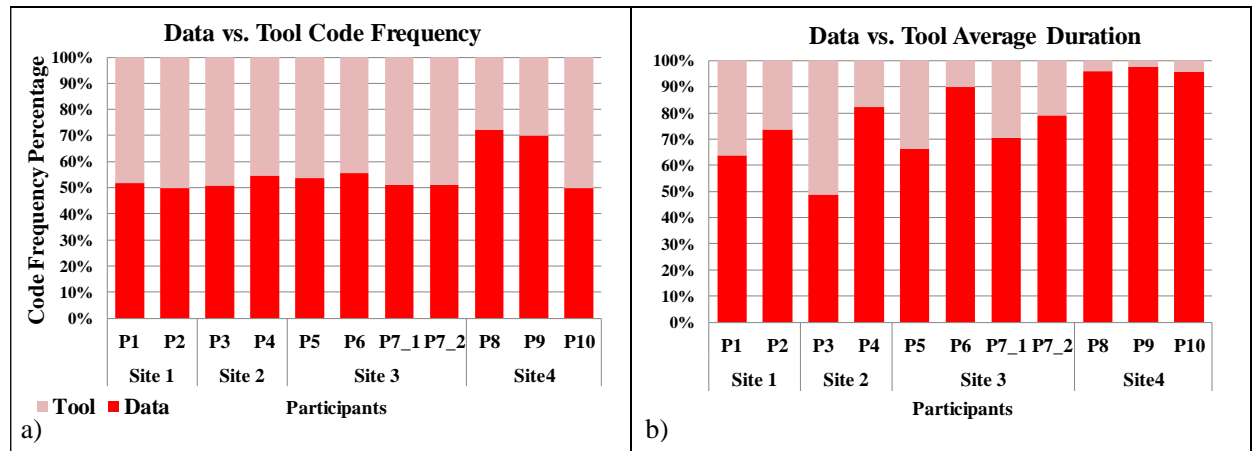

Figure. 10. Data vs. Tool: frequency and average duration.

**Boundary vs. Region Results:**

As shown in Figure 11 a, P1, P6 and P7 (Session 2) looked more at image feature boundaries while other participants looked more at regions. This was also reflected in the average duration (Figure 11, b). This suggests that different segmentation tools and datasets result in different gaze patterns and micro-decisions, varying the ratio of time spent looking at regions versus boundaries. P8-P10 spent more time on non-region of interests. Since these participants decided to segment the structures by marking (creating circles to capture structures), they spent more time on gazing at non-region of interests (close to the regions) to make sure they placed the circle correctly (in terms of size to cover the whole structure).

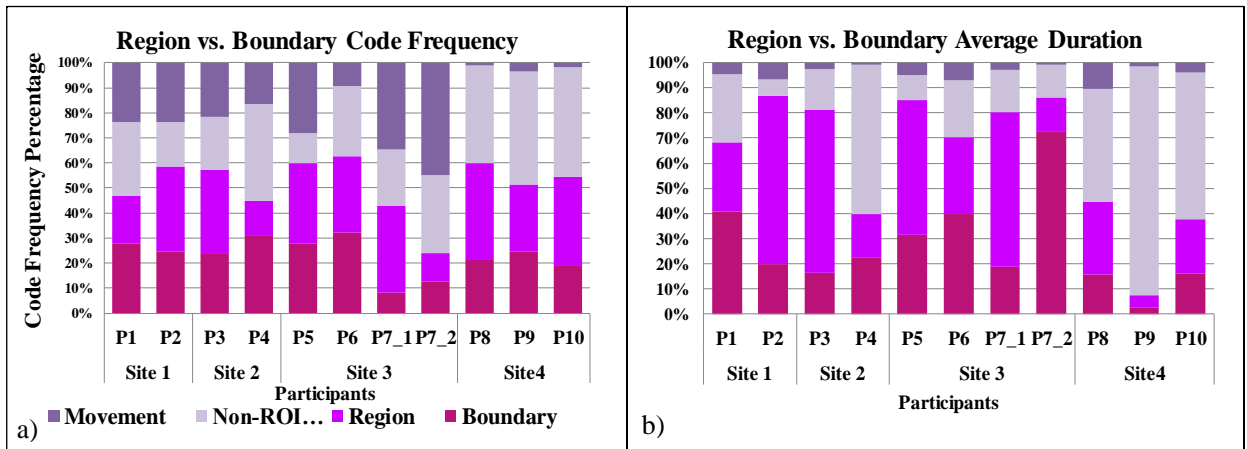

Figure. 11. Region vs. Boundary: frequency and average duration.

**Navigation Results:**

The participants showed no particular patterns for overall navigation (Figure 12). However, most participants (except P3) both paged more frequently — and spent more time paging— than any other navigation type.

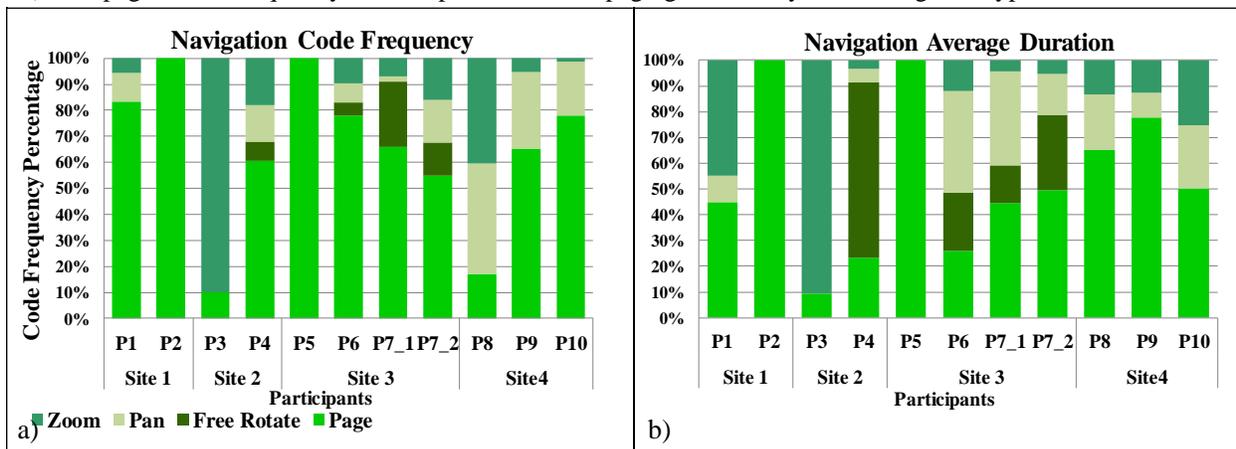

Figure. 12. Navigation: frequency and average duration.

**Marking Results:**

Regarding Marking micro-tasks, as shown in Figure 13, those participants who have higher code frequency for "Boundary" than "Region" also have higher code frequency for "Draw" compared to "Fill", and vice versa. P8-P10 spent more time on non-region of interests. Since these participants decided to segment the structures by marking (creating circles to capture the structures), they spent more time on gazing at non-region of interests (close to the regions). This shows a potential relation between different gaze patterns and Marking actions/decision. Those who have a boundary-based gaze pattern tend to do the segmentation by drawing while those with a region-based gaze pattern complete the segmentation process by filling the structure. Participants who segments the structure by marking (e.g., putting circles) tend to look more at non-ROIs. P3 did not take any marking action during the observation since they were just adjusting a slider.

**Review Results:**

According to Figure 14 there are code similarities on micro-decisions for P1 and P2 of site 1, P4 of site 2, P5 and P7 (Session 1) of site 3, and P8-P10 of site 4. P3 has a completely different code pattern (P3 was adjusting a slider). P7 (Session 2) has a higher code frequency for Access data. Both P5 and P6 spend more time on "Mark Review" comparing to P7.

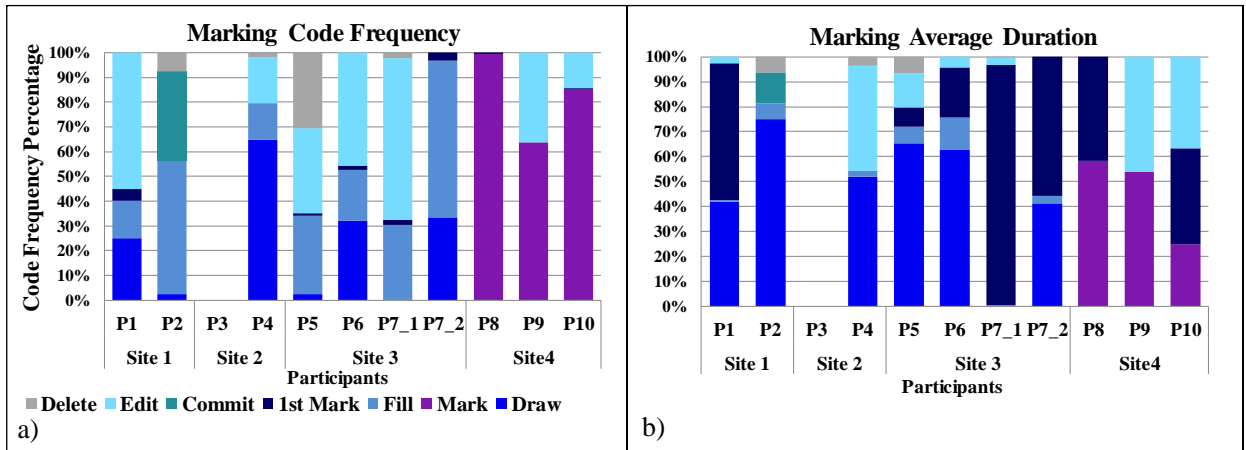

Figure. 13. Marking: frequency and average duration

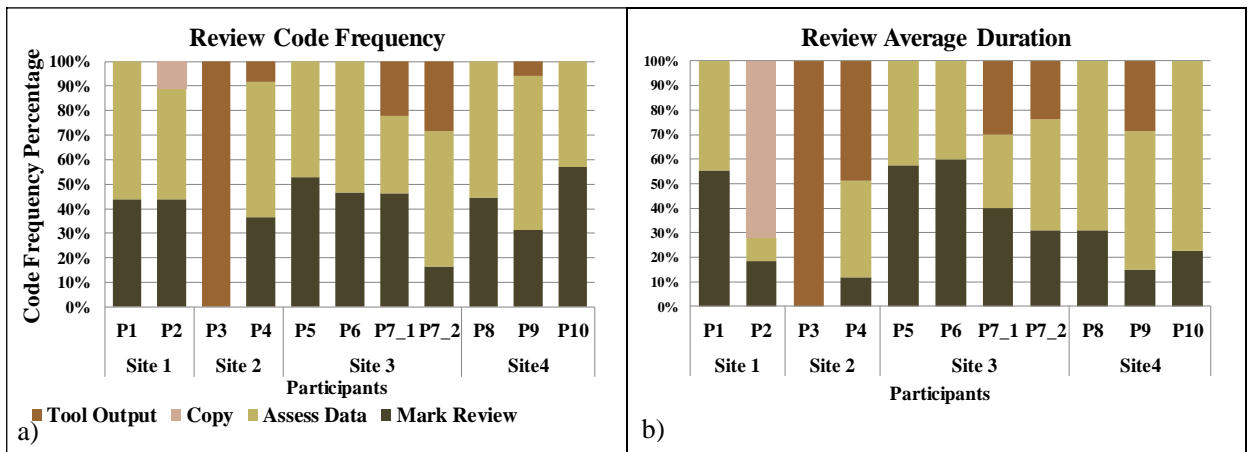

Figure. 14. Review: frequency and average duration.

## 5.2 Macro-Task Frequency Analysis and Results

Using macro-task categories, we have completed macro-task frequency analysis for 6 participants (P5-P10) whom we observed in the revised studies. For P7, we analyzed the two observations separately. P7 was an expert participant who segmented two different datasets in two different NO sessions. Similar to micro-decision analysis, we first count the frequency of each macro-task code over the entire valid observation/NO period. We then capture duration by taking the average time spent in each interval (e.g. if a participant has 20 "Review" task intervals, totaling 60 seconds, the average duration for "Review" macro-task is 60/20 = 3 seconds).

Figure 15 shows the category frequency and average duration for each of the participants. Different participants have different patterns of task frequency and average duration. The "Tool-usage" category has the highest frequency which indicates that participants mostly decide to work with the tools and use different mechanisms (e.g., drawing) to segment the structures. Although "Setup" frequency is low for all participants, the average duration indicates that "Setup" is a time intensive process for some of the participants (P8 and P10). One interesting difference between experts and novices is that the experts spend much more time viewing/working with the 3D structure (e.g., P7 and P10 are both experts and spend more time on 3D navigation).

## 5.3 Snapshot Analysis and Results

In this section, we analyze the data using video snapshots in order to distinguish how patterns of micro-decisions change over time for different tasks. We use tasks based on the participant's task-outline. For task snapshot analysis, we focused on four micro-decision eye-gaze location codes: "Boundary", "Region", "Non-ROI feature" and "Movement". We have snapshot analysis for the 6 participants whom we observed in our revised studies (P5-P10). We have two types of snapshot analysis:

1. Macro-task snapshot analysis: Count micro-decision eye-gaze codes (Boundary, Region, Non-ROI feature and Movement) frequency and average duration for each of the macro-task categories. We have macro-task snapshot analysis for 6 participants whom we observed in our revised studies (P5-P10).
2. Task-outline snapshot analysis: Count micro-decision eye-gaze codes (Boundary, Region, Non-ROI feature and Movement) frequency and average duration for selected task items of the modified task-outlines. This analysis helps us to determine what participants are doing during different segmentation tasks, and what their higher-level cognitive tasks and decisions are.

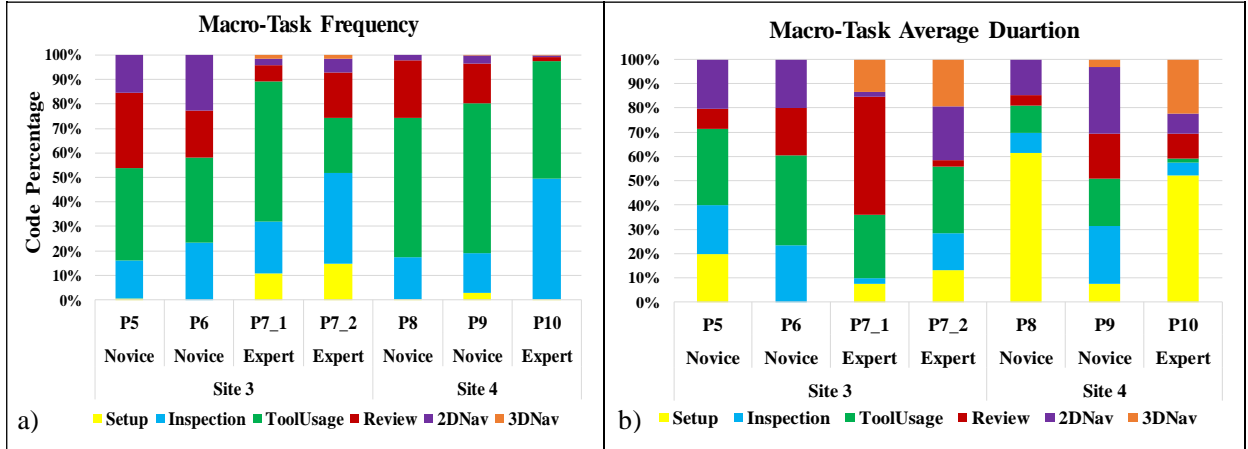

Figure. 15. Frequency and average duration for each macro-task.

### 5.3.1 Macro-Task Snapshot Analysis

We analyze each macro-task category (e.g., Setup) in detail by assigning one snapshot to each macro-task and then count the eye-gaze micro-decision frequency (Boundary, Region, Non-ROI feature and Movement) within each snapshot (1 count being 0.1 seconds). The final code percentage for each macro-task is separately computed by summing up all frequencies of the code in each snapshot. Figure 16 shows macro-task snapshot results (frequency and average duration) for site 3 (P5-P7) and site 4 (P8-P10). Please note we observed P7 in two sessions.

As shown in Figure 16 a) and b), similar to micro-decision analysis, P5 and P7 (Session 1) look more at regions and also spend more time looking at regions for almost all the macro-tasks (expect "Setup" for P5 which the participant spends more time looking at the boundary). In our micro-decision analysis, we showed that these participants decided to use the "Fill" approach to do the segmentation which involves looking more at regions. On the other hand, P6 and P7 (Session 2), who decided to use paint brush tool for segmentation, have higher code frequencies for boundary compared to region, and spend more time looking at boundaries. Comparing to novice users (P5 and P6), P7 (expert user) has a different pattern in terms of "Movement". Both P5 and P6 have fewer movements. At a high level, the expert paged back and forth between slices multiple times between fills/drawings, whereas the novice just filled or drew. One other clear difference between the expert and the novices is that the expert spent much more time up front evaluating the data as a whole before making a decision to begin marking.

Comparing to Site 3, participants of Site 4 have different patterns for frequency and average duration. Figure 16 c) and d) shows that during each of the macro-tasks participants of Site 4, who segment the structures by marking (e.g., putting circles) tend to look more at non-ROIs. We could not identify much difference between the expert and the novices of Site 4.

### 5.3.2 Task-Outline Snapshot Analysis

In the previous section, we completed snapshot analysis for each of the macro-tasks (e.g., how frequently does a participant look at boundaries during the "Review" macro-task). But we are also interested to know what participants are doing during different tasks and what their higher-level decision elements are (please note by task we mean a task item included by the participant in the modified task-outlines). Our goal is to identify if there is any pattern between different/similar tasks/decision elements for different participants, and how similar the gaze patterns are for repeated instances of one task example.

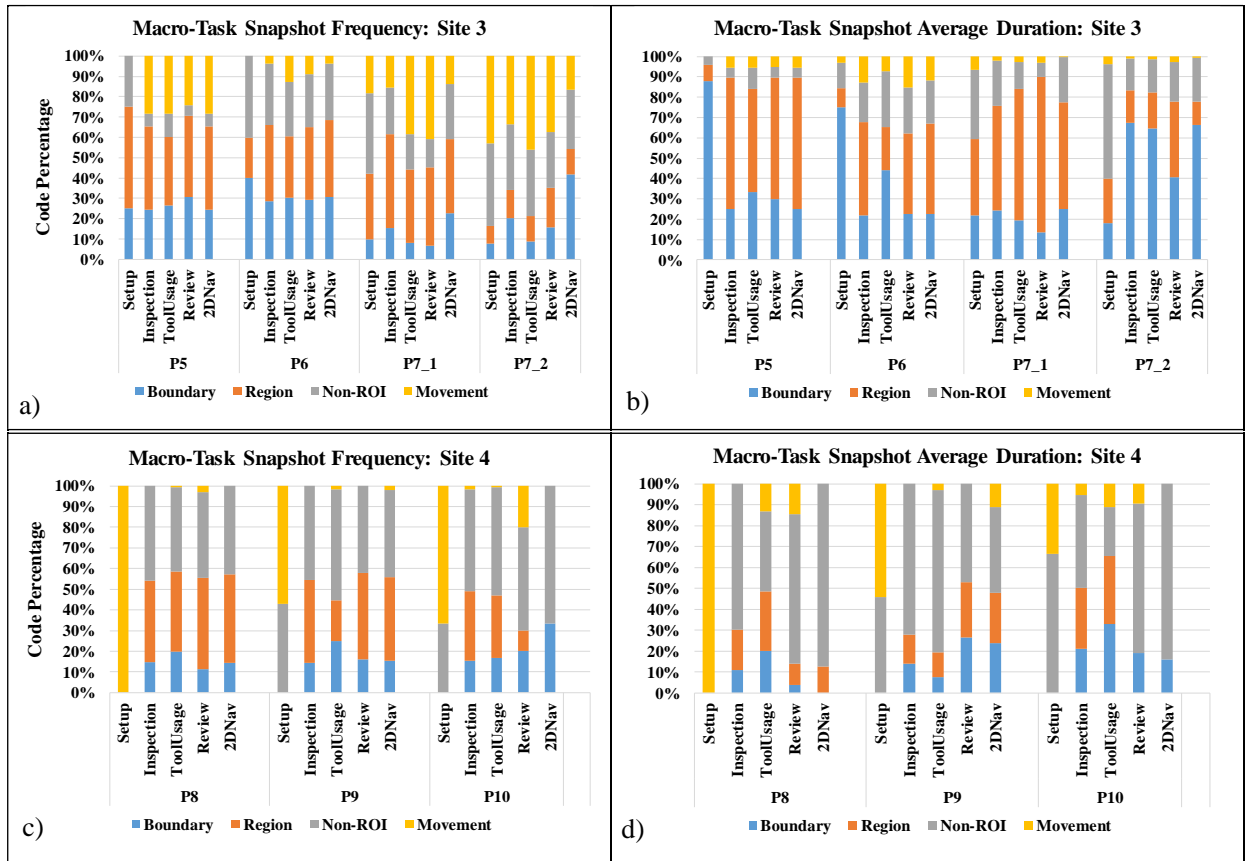

Figure. 16. Macro-task snapshot analysis: Frequency and average duration for participants: a) and b) Site 3 (P5-P7); c) and d) Site 4 (P8-P10).

For task-outline snapshot analysis we picked two participants from each site (one expert and one novice). We also picked two different individual tasks/decision elements for each of the participants. Table 5 summarizes the participants and tasks for task-outline snapshot analysis. We chose the task examples in a way to insure similarities in task types. For instance, for P6 and P7, task examples 1 are similar because both tasks include tracing boundaries and using a drawing (brush/lasso) tool. Task examples 2 are also similar as they both include flipping between slices.

Table 5. Participants' Task Examples

| Participant | Status | Task/Decision Element Example 1 | Task/ Decision Element Example 2 |
|---|---|---|---|
| P6 | Novice | Use Paint Brush Tool | Flip Between Slices |
| P7 Session 2 | Expert | Trace ROI boundary and use Lasso tool | Flip between slices |
| P8 | Novice | Identify Cell | Drop blue disks to fill cell |
| P10 | Expert | Identify Cell | Drop blue disks to fill cell |

**Results for P6 and P7 (Session 2):** For P6 who is a novice participant at Site 3, we chose the two individual sample tasks/decision elements "Example 1: Use Paint Brush Tool" and "Example 2: Flip Between Slices". We captured 9 repeated task instances of P6 task example 1 (T1-T9), and 8 instances of task example 2 (T1-T8). For the expert segmenter (P7 Session 2), we chose the two individual sample tasks "Example 1: Trace ROI boundary and use Lasso tool", and "Example 2: Flip Between Slices". We captured 6 repeated task instances of task example 1 (T1-T6), and 7 instances of task example 2 (T1-T7).

Figures 17 shows the analysis results for P6 and P7 Task example 1. P6 decides to use the paint brush tool for annotation to capture the boundaries. P7 uses the lasso tool to trace the boundaries. Our previous results showed that participants who decide to use drawing for segmentation, also look more at boundaries (boundary-based approach).

Therefore, we expect P6 and P7 (session 2) to look more at boundaries while doing the annotations. Results of analysis confirm this. Both P6 and P7 look more at boundaries and spend more time on boundaries during task example 1.

Figures 18 shows the analysis results for P6 and P7 Task/decision element example 2. Although the task is similar for both participants (Flip between slices), P6 and P7 have different gaze patterns for the task. The duration for "Movement" is higher for P6. P7 still spends more time looking at boundaries and has lower average duration for movement. Both participants have higher code frequency and average duration for "Non-ROI" code in their task example 2 (compared to task example 1).

P6 and P7 (Session 2) have a boundary-based strategy of segmentation and look/spend more time on boundaries. However, they have different task-outlines and gaze patterns even if they worked with similar tools and data sets. In addition, gaze patterns were different in different individual tasks, but repeated instances of tasks show more similar patterns.

**Results for P8 and P10:** For P8 (novice) and P10 (expert), we chose the two individual sample tasks/decision elements "Example 1: Identify Cell" and "Example 2: Drop blue disks to fill cell". For both participants, we captured 10 repeated instances of task example 1 (T1-T10). For task example 2, P8 has 9 instances (T1-T9), and P10 has 12 instances (T1-T12). Figures 19 shows the analysis results for P8 and P10 Task example 1. They both look more at non-ROI features. However, P8 almost does not look at boundaries, while P10 looks more and spends more time on boundaries.

Figure 20 shows the analysis results for P8 and P10 Task example 2. Both participants use marking mechanisms (dropping circles) for annotation. Although the task is similar for both participants, but they have different gaze patterns for the task. While P10 almost does not look at regions, P8 looks more and spends more time on boundaries. Both participants have higher frequency for "Non-ROI" code.

Again, as we can see in figures, P8 and P10 have non-region based strategy of segmentation and look/spend more time on "Non-ROIs". They have different task-outlines and gaze patterns even if they worked with similar tools and data sets and cover similar tasks. In addition, gaze patterns were different in different individual tasks, but repeated instances of tasks show more similar patterns for each participant.

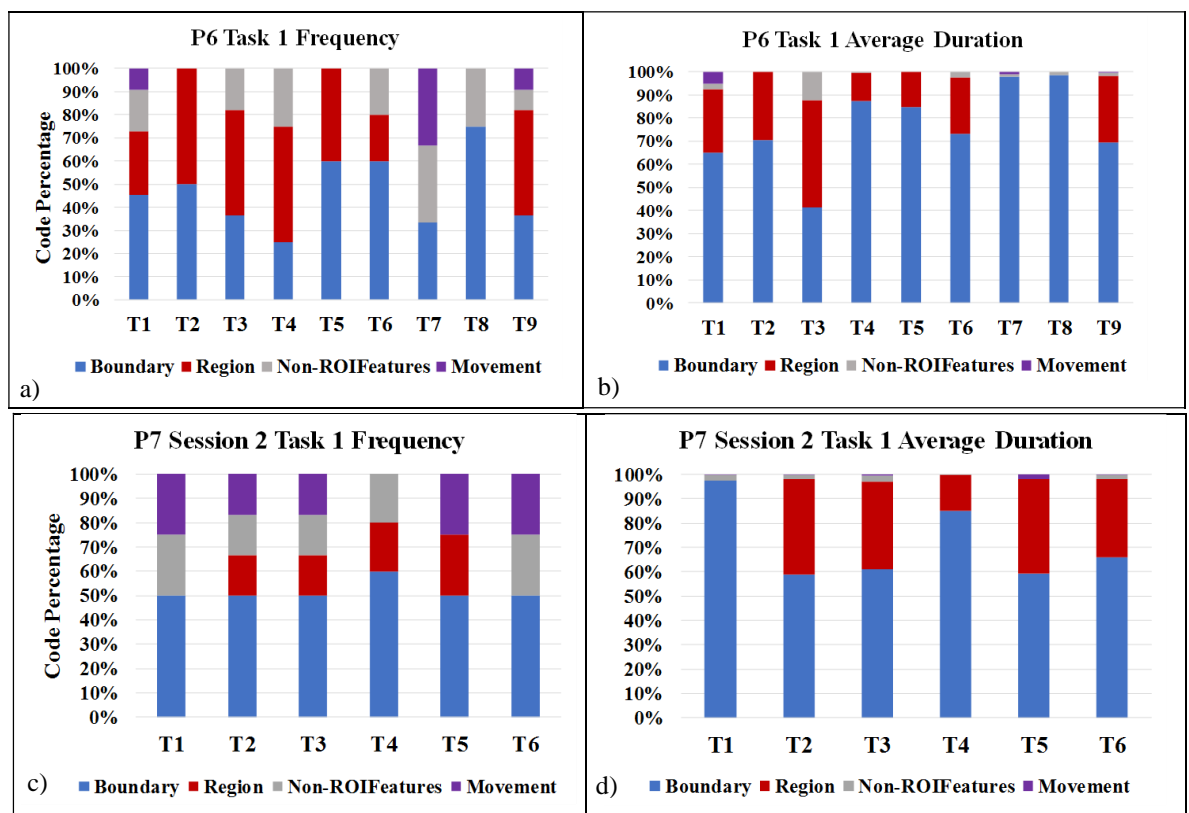

Figure. 17. Code frequency and average duration for Task Example 1. a) and b) P6, Use Paint Brush Tool; c) and d) P7, Trace boundary and use lasso tool.

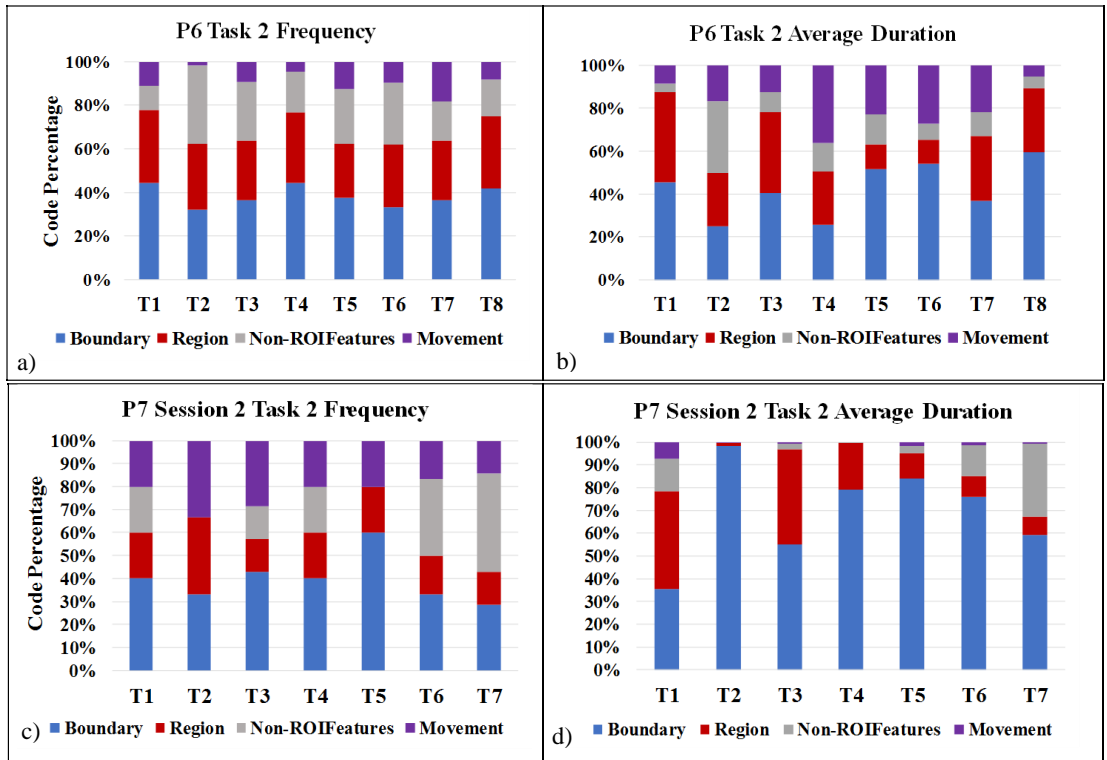

Figure. 18. Code frequency and average duration for Task Example 2 (Flip Between Slices). a) and b) P6; c) and d) P7

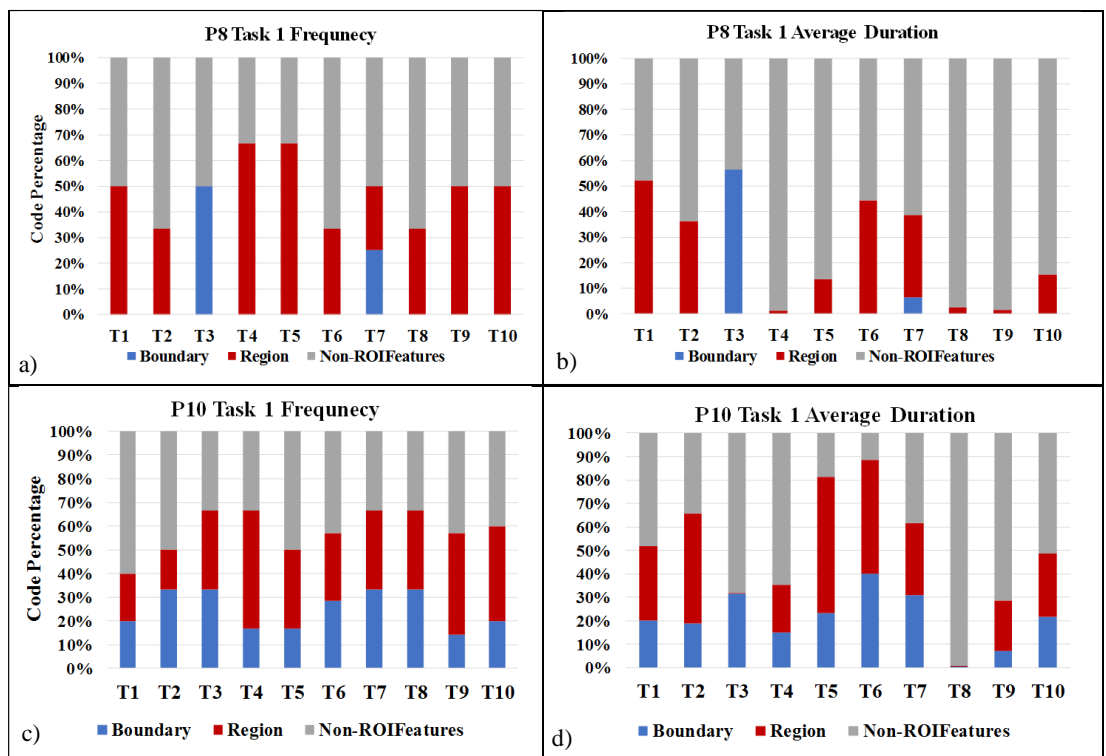

Figure. 19. Code frequency and average duration for Task Example 1, Identify Cell. a) and b) P8; c) and d) P10

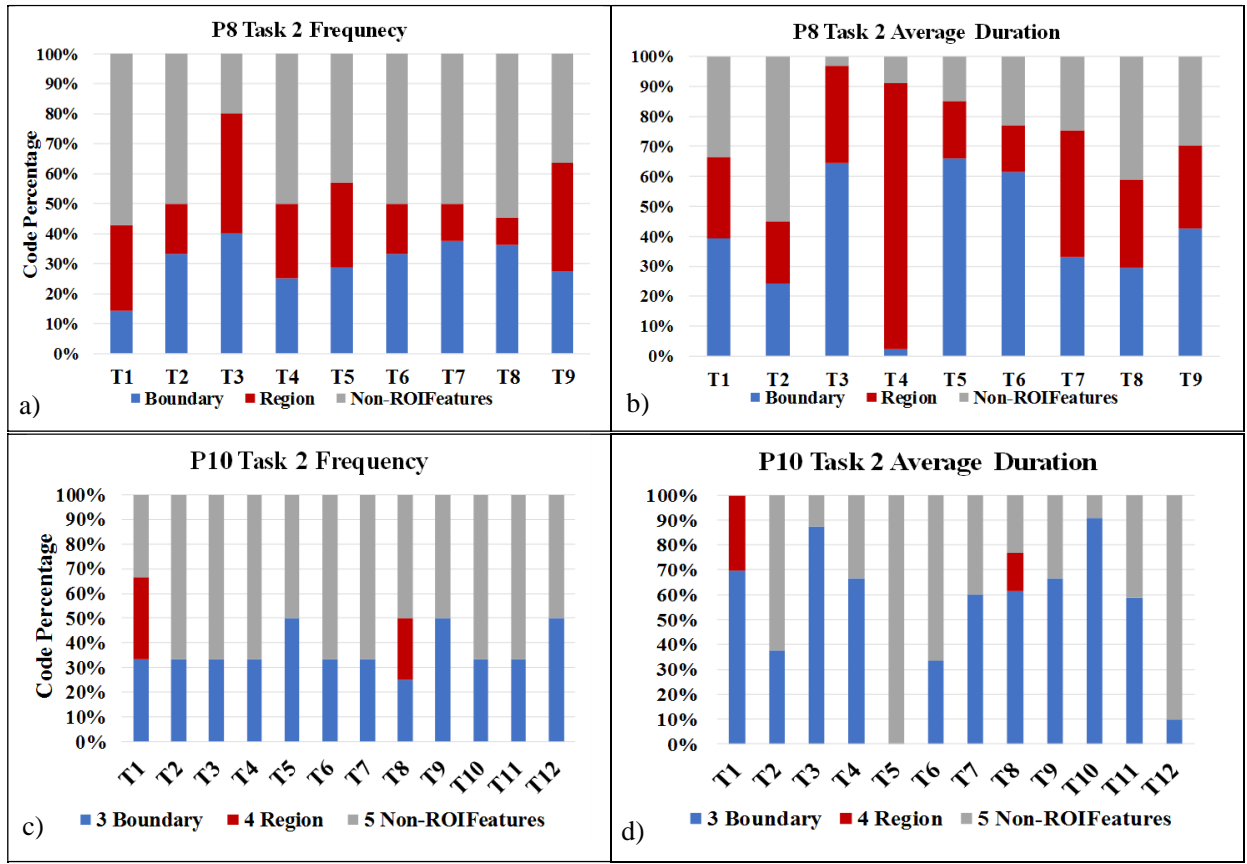

Figure. 20. Code frequency and average duration for Task Example 2, Drop blue disks to fill cell. a) and b) P8; c) and d) P10.

## 6. Summary of Results and Discussion

**RQ1:** Using our micro-decision coding scheme and macro-task classification, we could successfully quantify where segmenters look, and what their low-level decision elements and higher-level tasks are. We identified that depending on the segmentation tool and the data set, participants utilized one of these three segmentation decision strategies: Region-based (seeing data as regions to be filled/ captured), boundary-based (seeing data as boundaries to be demarcated by drawing contours), and non-ROI-based. Additionally, if participants explicitly captured a structure by drawing, they looked more at the boundaries, filling resulted in more gazing at regions, and marking (e.g., putting circles) involved looking more at non-ROIs. Table 6 shows participants' segmentation decisions strategies and approaches.

For region and boundary-based strategies, participants exhibited approximately equal amounts of gaze movement between the data and the tool. For non-ROI based strategies, participant looked more at data. All participants, on average, spent more time gazing at data than tools.

**RQ2:** Different participants had different task-outlines and gaze patterns even if they worked with similar tools and data sets. One participant utilized two different segmentation approaches (same data set and tools, but a different structure). In addition, gaze patterns were different in individual task examples, but repeated instance of the same task had similar patterns.

Table 6. Participants' Marking methods and segmentation approaches

| Participant | Marking Method | Segmentation Decision Strategy |
|---|---|---|
| P1 | Draw Contours | Boundary-based |
| P2 | Fill Structure | Region-based |
| P3 | Adjust Slider | Region-based |
| P4 | Draw Contours | Boundary-based |
| P5 | Fill Structure | Region-based |
| P6 | Draw | Boundary-based |
| P7 (Session 1) | Fill Structure | Region-based |
| P7 (Session 2) | Draw | Boundary-based |
| P8 | Put Circle Marks | Non-ROI-based |
| P9 | Put Circle Marks | Non-ROI-based |
| P10 | Put Circle Marks | Non-ROI-based |

Also, in general, comparing to novices, experts spent more time working with 3D structures/views. According to these results, we hypothesize that given a 3D structure and slicing plane, experts can: 1) predict the 2D contour; 2) predict how 2D contour changes with small view changes; and 3) identify invalid 2D contours. To test our hypothesis, we are developing/validating a test instrument which measures participants' spatial ability while dealing with 2D slices of 3D structures (Sanandaji, Grimm, & West, 2016, 2017).

## 7. Conclusions

We developed a hybrid field-study protocol that incorporates elements of CTA, CDM, and eye-tracking/marking data to capture visual-spatial decision-making process of a 3D volume segmentation task. We presented a novel taxonomy for coding observations of 3D segmentation process and used the coding scheme to successfully capture and analyze 10 segmenters' perceptual cues, low-level decision-making elements (micro-decisions). We used the coded data to identify three potential segmentation decision strategies (region, boundary, and non-ROI-based). We also introduced our macro-task classification mechanism and observed how gaze and task patterns change over time for different segmentation strategies. In addition, we compared novices with experts in terms of both micro-decisions and macro-tasks to capture their differences in terms of tasks and decisions. For example, one key difference between experts and novices is that the experts spend much more time viewing/working with the 3D structure. We could identify where experts look and what their low-level (micro) decisions and higher-level (macro) segmentation actions and tasks are.

To ensure reliable study results with this small sample size, we focused on rigorous data collection from multiple sources (video, audio, eye-gaze) and cross-validation between the different data streams. Our focus was not on statistical sample analysis but rather on a repeatable coding scheme and task-flow visualization to identify particular patterns both within and between participants. With our analysis, we effectively captured, in a quantitative manner, the tasks and behaviors observed qualitatively. We are using these results in other parts of our research, with the ultimate goal of creating training guidelines to improve segmentation process.